\documentstyle[aaspp4,12pt]{article}   

\slugcomment{\small Submitted to {\sl ApJL\/}, 19 Nov 1997, 
accepted 13 Jan 1998}
\lefthead{S. Levine \& L. Sparke}
\righthead{A model for lopsided galactic disks}

\def\eg{{e.g.}}
\def\etal{{et al.}}
\def\kms{\mbox{\,km~s$^{-1}$}}

\begin{document}

\title{A model for lopsided galactic disks}
\author{Stephen E. Levine}
\affil{Observatorio Astron\'omico Nacional, IA-UNAM \\
Ensenada, B.C., M\'exico \\ 
and \\
U. S. Naval Observatory, Flagstaff Station \altaffilmark{1} \\
P.O. Box 1149, Flagstaff, AZ 86002-1149, USA}
\authoraddr{U.S. Naval Observatory, Flagstaff Station, P.O. Box 1149
Flagstaff, AZ 86002-1149, USA}
\altaffiltext{1}{Current address}
\and
\author{Linda S. Sparke}
\affil{Washburn Observatory, 475 North Charter Street \\
Madison WI 53706-1582, USA}

\begin{abstract}

Many disk galaxies are lopsided: their brightest inner parts are
displaced from the center of the outer isophotes, or the outer
contours of the H{\sc i} disk.  This asymmetry is particularly common
in small, low-luminosity galaxies.  We argue here that long-lived
lopsidedness is a consequence of the disk lying off-center in the
potential of the galaxy's extended dark halo, and spinning in a sense
retrograde to its orbit about the halo center.  The stellar velocity
field predicted by our gravitational $N$-body simulations is clearly
asymmetric.

\end{abstract}

\keywords{Galaxies: Evolution -- Galaxies: Irregular --
Galaxies: Kinematics and Dynamics -- Galaxies: Structure}

\section{Introduction}

Many galaxies display lopsided, rather than bisymmetric, structure, on
both large and small scales.  Hubble Space Telescope observations show
that the nuclei of M31 and NGC 4486B do not lie at the center of the
bulge isophotes (Lauer \etal\ 1993, 1996; Davidge \etal\ 1997).  In
our own Milky Way, the distribution of molecular gas as measured in CO
is offset about $1^{\circ}$ in longitude and +15\kms\ in velocity from the
center (Binney \etal\ 1991); other mass concentrations in that region
also appear to be off-center (Blitz 1995).  On kiloparsec scales, many
galaxies are asymmetric in both their optical appearance and their gas
distribution (\eg\ Baldwin, Lynden-Bell \& Sancisi 1980).  Working
from a sample of 1700 spectra, Richter \& Sancisi (1994) estimate that
about half of all late-type spiral galaxies show clearly asymmetric
H{\sc i} profiles, indicating a lopsided gas disk; about half of this
sample are nearby field galaxies, and none lie in rich clusters where
interactions between galaxies are frequent.  Zaritsky \& Rix (1997)
estimate that about 50\% of spiral galaxies in the field are
significantly lopsided at optical and near-infrared wavelengths; the
asymmetry involves both young and old stellar populations, and most of
their lopsided galaxies lack any obvious companions.  Asymmetric bars
are particularly common in late type or small disk galaxies, less
luminous than $M_B \sim 18$ (\eg\ Feitzinger 1980; Odewahn 1996;
Matthews \& Gallagher 1997).  
While some lopsidedness in the outer parts of galaxies is undoubtedly
due to recent accretion, the fact that lopsided asymmetries are so
common suggests that they can be long-lived.

The origin and persistence of these disk asymmetries remain a
mystery.  Baldwin, \etal\ (1980) suggest that the lopsided distortions
seen in the outer parts of galaxies are kinematic features; in that
case they should wind themselves into a leading spiral, since the slow
pattern speed $\Omega - \kappa$ is negative. In fact the one-armed
spiral often has the same sense of winding as the two-armed
components, which presumably trail (Colin \& Athanassoula 1989;
Phookun \etal\ 1992, 1993).  The only analytic disk models which are
strongly unstable to lopsided distortions are those with many
retrograde-streaming particles (Zang \& Hohl 1978; Sawamura 1988;
Sellwood \& Merritt 1994); but observed counter-rotation is rare in
disk galaxies (\eg\ Kuijken, Fisher \& Merrifield 1996).  Adams, Ruden
\& Shu (1989) suggested that a protostellar disk could become unstable
to a lopsided distortion; in the galactic context, this would
correspond to the stellar disk being off-center in the dark halo.
However, Heemskerk, Papaloizou \& Savonije (1992) showed that the
$m=1$ distortion is generally evanescent in the disk, so that resonant
growth will not occur.  Matthias (1993), investigating orbits in a
weakly `egg-shaped' tumbling potential, found that prograde orbits
were distorted so as to oppose the `eggness'.  Miller \& Smith (1992)
describe an $N$-body simulation of an elliptical galaxy in which a
central tilted disk 
developed of an off-center motion, tracing an oscillation in the
underlying particle distribution,
and Taga \& Iye (1998) have done simulations in which a massive object
wanders off-center in a dense stellar system;
but no similar results have been reported in the literature.  
By contrast, kinematic models for lopsided systems have been quite
successful; orbits in the potential of an off-center bar rotating
rigidly in an axisymmetric disk can account for the rotation curves of
Magellanic barred systems (de Vaucouleurs \& Freeman 1973;
Christiansen \& Jeffreys 1976), while the gas response has a trailing
one-armed form (Colin \& Athanassoula 1989).

We propose that the key to the 
puzzling ubiquity of lopsided galactic disks is
that they are not dynamically isolated inside the galaxy.
Late-type spirals 
and dwarf galaxies, in which larger-scale lopsidedness is most
frequently observed, are most likely to be dominated by dark halo
mass: \eg\ Casertano \& van Gorkom (1991); C\^ot\'e, Carignan \&
Sancisi (1991); Broeils (1992a, b).  (Athanassoula, Bosma \& Papaoiannou
1987 suggested that in later-type galaxies the halo core is larger in
relation to the radius of the stellar disk, although this was not
confirmed by Broeils 1992a, b.)  If a disk found itself off-center in a
dominant dark halo with a core of nearly constant density that was
large compared to the disk dimensions, most of the disk mass would lie
in a region where the angular speed of an orbit about the halo center
did not vary strongly.  Self-gravity might then act to maintain the
disk's coherence as it orbits the center of the halo potential well.
This paper presents our investigation of just such a model using a
self-consistent, self-gravitating disk embedded in a static dark halo.

\section{Gravitational $N$-body simulations and results}

We used a tree code (\eg\ Barnes \& Hut 1989) implemented in NEMO
(Teuben 1995) to follow the development of a rotating disk which was
set up in equilibrium at the center of a fixed `halo' potential, then
displaced off-center; every disk particle was also given a `sideways'
velocity appropriate to a circular orbit in the halo potential at the
radius of the disk center.  Our disk was represented by $20 \, 000$
particles, distributed in a Kuzmin-Toomre disk (eq. 2-49 of Binney \&
Tremaine 1987) of unit scale length, which we truncated at $r=10$.
The velocity dispersion in the disk plane was set so that Toomre's
(1964) stability parameter $Q=1.5$; for our 3-D simulations, we then
`puffed up' the disk, by setting the vertical velocity dispersion
equal to half that in the radial direction.  The halo was of
pseudo-isothermal form:
\begin{equation}
\rho(r) = { {\rho_0} \over {1 + r^2/r^2_c} } \ .
\end{equation}
The density $\rho (r)$ is approximately constant within the core
radius $r_c = 2$; the speed of a circular orbit in this potential
rises linearly in the core, becoming flat at large radii.  We took
$G=1$, and chose $\rho_0$ and the disk density so that the total mass
inside $r=20$ is unity. The halo contributes 90\% of this total, and
accounts for 60\% of the mass within 5 disk scale lengths of the
center, consistent with the findings of Broeils (1992a, b) for galaxies
with $V_{\rm rot} \lesssim 120\kms$.  The time step $\Delta t$ was $0.1$,
and the explicit softening parameter $\epsilon = 0.05$, close to the
optimum suggested by the method of Merritt (1996); moderate variations
in $\epsilon$ made little difference.  The opening angle for the
tree code was set at $0.75$, providing a good balance between
resolution and speed.

When the disk spin is retrograde with respect to its orbit around the
halo center, we find that the disk remains off-center while making several
orbits of the halo. Figure~1 shows a run in which the disk center was
initially displaced by $x_0 = 2.5$, so that about 20\% of its mass lay
at $r_c \lesssim 2$, within the halo core.  The velocity dispersion is
sufficiently low that the disk formed a bar, which persisted
throughout the simulation.  Freeman (see de Vaucouleurs \& Freeman
1973) proposed a model for Magellanic barred galaxies in which the bar
was displaced parallel to its minor axis from the center of a rotating
disk; calculations of periodic closed orbits in Freeman's potential
showed that the model could reproduce the observed offset between the
center of the rotation curve measured in the gas and the bar center.
The lower section of Figure~1 shows the `stellar' rotation curve of
our model galaxy, as it would be seen by an observer for whom the
minor axis of the bar lies in the plane of the sky; the distance $\xi$
is measured from the centroid of the most-bound bar particles.  The
center of rotation, at $v_\perp = 0$, is displaced from that centroid,
and the rotation curve is clearly asymmetric.

When started with a larger displacement, $x_0 =5$, our retrograde disk
drifted inwards to $r \gtrsim 2$, just outside the halo core (see
Figure~2).  If we gave the disk insufficient angular momentum for a
circular orbit in the halo, it first moved inwards, and then onto a
near-circular orbit within the core.  The lower portion of Figure~2 is
for a disk with prograde rotation, which drifts inwards more strongly
than one in retrograde orbit; even so, it stays at $r \gtrsim 1$ while
making several orbits of the halo.  In runs with $r_c = 0.5$, where
the core radius of the halo is not large compared with the scale
length of the disk, neither the prograde nor the retrograde disk
remained far off-center.

\section{Discussion}

Our experiments imply that a gravitating galactic disk can remain
off-center in a halo potential as long as it orbits in a region where
the halo has nearly constant density; the effect is more pronounced if
the spin of the disk is in the opposite sense to its orbit around the
halo center.  This behavior can be understood in the limit where the
disk dimensions are small, and its self-gravity is weak: a particle in
near-circular orbit in the halo then follows an epicycle in a sense
retrograde to the orbit.  A collection of particles with the same
angular momentum, and hence with a common guiding-center radius, would
stay together as they orbited the halo center.  In the halo core,
where the orbital frequency does not vary strongly with radius, the
disk's self-gravity can fairly easily maintain coherence in
approximately this configuration.

The relatively large fraction of lopsided systems found at redshifts
$z \sim 1$, which was apparently a period of rapid galaxy formation
(\eg\ Driver, Windhorst \& Griffiths 1995), suggests that many disks
can slowly lose their off-center character as they settle down.  In
dwarf galaxies, the halo is relatively more massive (Broeils 1992a, b), so
lopsidedness is more easily maintained than in more luminous systems.
The relation between a lopsided disk and the presence of a central bar
has not been studied systematically, but both these features are
characteristic of Magellanic irregular systems.  The lopsided Sc
galaxy NGC~1637 (Block \etal\ 1994) appears prominently barred in an
infrared image, while at optical wavelengths the bar is hidden by dust
and confused by star formation.  The spatial scale of the predicted
asymmetry is approximately that of the constant-density core;
consistent with this idea, the off-center nuclei of M31 and the
elliptical NGC~4486B lie within resolved central cores in those
galaxies.

Our results are consistent with those of Bontekoe (1988), who
investigated the sinking of a satellite galaxy into a larger system.
In his unpublished fifth chapter, he reports that the satellite never
sank all the way to the center, but the orbit shrank until it enclosed
about 10\% of the main galaxy's mass.  
The larger system was represented by a series of polytropes; 
the more concentrated the polytrope, the closer the satellite sank 
to its center.  Our study also
has aspects in common with work on galaxy mergers by
Miller \& Smith (1995), who found that if two polytropic `cores'
orbited each other in isolation, dynamical friction caused them to
merge within a couple of orbits.  But in the presence of a `halo' of
constant density, whether it was represented by a fixed potential, or
consisted of `live' simulation particles, the cores survived for
$15-30$ orbits without coming noticeably closer, although the
interpretation of the `live halo' model was complicated by an
overstable oscillation in the cores' orbital radii.  Persistent
off-center disks have not been reported in published $N$-body
simulations following a galactic disk in an imposed external bulge
potential (see \eg\ Sellwood \& Wilkinson 1993); but in these
computations, the mass distribution of the bulge was in general more
concentrated than the disk.

We must be concerned about the effect of dynamical friction on the
disk, from the particles which would make up a `live' galactic halo.
In the context of M31's nucleus, King, Stanford \& Crane (1995) point
out that this drag could well be small, if the bulge stars stream
rapidly in the same direction as the orbiting cluster's motion.  
It is unclear even how far the usual estimates of dynamical friction
are to be trusted.
Bontekoe's (1988) satellite stopped sinking at some distance from the
center of the system into which it was accreting. 
The physical situation of an off-center disk has much in
common with that of a rotating bar, where estimates of the drag from
the particles of a spherical halo suggest that the bar should rapidly
spin down (Weinberg 1985), and gravitational $N$-body experiments also
indicate strong braking.  
But observed galactic bars seem to be
fast-rotating: Sellwood (1996) summarizes this confusing situation.
The long-lived pairs of {\it counter-rotating} bars which develop in
the gravitational $N$-body simulations of Sellwood \& Merritt (1994)
and Friedli (1996) are clearly not discouraged by dynamical friction.

It may well be more useful, instead of calculating only the
back-reaction of the halo on the orbiting disk, to look for long-lived
modes of oscillation in the combined system.  Weinberg (1991, 1994)
has developed an analytic method for doing this; examining spherical
King models with an isotropic velocity dispersion, he found $m=1$
modes which, once excited, require many orbital times to decay.  The
distortions lasted longest in the models in which the core was largest
in relation to the total extent of the system.  He then used
gravitational $N$-body simulations to confirm the existence of these
modes.  Similar lopsided modes may well exist for a disk within a
galactic halo.

We would then expect a relatively long-lived lopsided structure 
to result when material is accreted such as to push it off-center in
the underlying potential.  
In the central regions of a galaxy, where
the gravitational force is provided largely by a `hot' stellar system
such as a bulge, an inner disk or nuclear star cluster could remain
off-center over many orbital periods.  Gas might be captured by 
a galactic disk such as to push it
off-center in the halo; alternatively, accretion of dark matter onto
the halo may result in the disk lying off-center.  If the orbit of the
disk is retrograde with respect to its spin, the lopsidedness is
likely to be stronger and more persistent.  To maintain appreciable
asymmetry, the near-constant-density halo core should be large enough
to encompass a substantial fraction of the disk mass.

\acknowledgments 
We are grateful to Dr. R.H. Miller for a helpful referee report, for
being willing to communicate with us directly,
and for reminding us of Bontekoe's work.
LSS thanks UNAM's Observatorio Astron\'omico Nacional
at Ensenada (BC, M\'exico) and USNO at Flagstaff for hospitality during
the course of this work, and the National Science Foundation for
funding through grant AST--93220403.  SEL thanks the Astronomy
Department of the University of Wisconsin-Madison for hospitality, and
the Consejo Nacional de Ciencia y Tecnolog\'{\i}a of M\'exico for
funding under grant 3739-E.

\pagebreak

\centerline{\bf References}
{
\setlength {\parskip}{0.0truecm}
\setlength {\parindent}{-0.4truecm}

Adams, F.C., Ruden, S.P. \& Shu, F.H. 1989, \apj, 347, 959

Athanassoula, E., Bosma, A. \& Papaioannou, S. 1987, \aap, 179, 33

Baldwin, J.E., Lynden-Bell, D. \& Sancisi, R. 1980, \mnras, 193, 313

Barnes, J.E. \& Hut, P. 1989, \apjs, 70, 389

Binney, J.J., Gerhard, O.E., Stark, A.A., Bally, J. \& Uchida, K. 1991, 
\mnras, 252, 210

Binney, J.  \& Tremaine, S. 1987, `Galactic Dynamics' (Princeton
University Press)

Blitz, L. 1995, in `The Physics of Gaseous and Stellar Disks of Galaxies',
ed. I.R. King, ASP Conf. Ser., 66, 1

Block, D.L., Bertin, G., Stockton, A., Grosb{\o}l, P., \& Moorwood, A.F.M. 
1994, \aap, 288, 365

Bontekoe, T.R. 1988, PhD Thesis, Groningen University, Netherlands

Broeils, A.H. 1992a, PhD Thesis, Groningen University, Netherlands 

Broeils, A.H. 1992b, \aap, 256, 19

Casertano, S. \& van Gorkom, J.H. 1991, \aj, 101, 1231

Christiansen, J.H. \& Jeffreys, W.H. 1976, \apj, 205, 52

Colin, J., \& Athanassoula, E. 1989, \aap, 214, 99

C\^ot\'e, S., Carignan, C. \& Sancisi, R. 1991, \aj, 102, 904

Davidge, T.J., Rigaut, F., Doyon, R. \& Crampton, D. 1997, \aj, 113, 2094

de Vaucouleurs, G. \& Freeman, K.C. 1973, Vistas in Astronomy, 14, 163

Driver, S.P., Windhorst, R.A. \& Griffiths, R.E. 1995, \apj, 453, 48

Feitzinger, J.V. 1980, Space Science Reviews, 27, 35

Friedli, D. 1996, \aap, 312, 761

Heemskerk, M.H.M., Papaloizou, J. \& Savonije, G.J. 1992, \aap, 260, 161

King, I.R., Stanford, S.A. \& Crane, P. 1995, \aj, 109, 164

Kuijken, K., Fisher, D. \& Merrifield, M.R. 1996, \mnras, 283, 543

Lauer, T.R. \etal\ 1993, \aj, 106, 1436

Lauer, T.R. \etal\ 1996, \apjl, 471, L79

Matthews, L. \& Gallagher, J.S. 1997, \aj, 114, 1899

Matthias, M. 1993, Diplomarbeit, Ruprecht-Karls--University, Heidelberg

Merritt, D. 1996, \aj, 111, 2462

Miller, R.H. \& Smith, B.F. 1992, \apj, 393, 508

Miller, R.H. \& Smith, B.F. 1995, unpublished manuscript

Odewahn, S. 1996, in IAU Colloq. 157, Barred Galaxies, eds. R. Buta,
D. A. Crocker \& B. G. Elmegreen (ASP Conf. Ser. 91)(San Francisco: ASP), 30

Phookun, B., Mundy, L.G., Teuben, P. \& Wainscoat, R.J. 1992, \apj, 400, 516

Phookun, B., Vogel, S.N. \& Mundy, L. 1993, \apj, 418, 113 

Richter, O.-G. \& Sancisi, R. 1994, \aap, 290, L9

Sawamura, M. 1988, \pasj, 40, 279

Sellwood, J.A. 1996, in IAU Colloq. 157, Barred Galaxies,
eds. R. Buta, D. A. Crocker \& B. G. Elmegreen (ASP Conf. Ser. 91)(San
Francisco: ASP), 259

Sellwood, J.A. \& Merritt, D. 1994, \apj, 425, 530

Sellwood, J.A. \& Wilkinson, A. 1993, Rep. Prog. Phys., 56, 173

Taga, M. \& Iye, M. 1998, \mnras, submitted

Teuben, P.J. 1995, in ADASS IV, ed. R. Shaw \etal, ASP Conf. Ser. 77,
398

Toomre, A. 1964, \apj, 139, 1217

Weinberg, M.D. 1985, \apj, 213, 451

Weinberg, M.D. 1991, \apj, 368, 66

Weinberg, M.D. 1994, \apj, 421, 481

Zang, T.A. \& Hohl, F. 1978, \apj, 226, 521

Zaritsky, D. \& Rix, H.-W. 1997, \apj, 477, 118

}

\pagebreak

\noindent

\begin{figure}
\hbox to \hsize{\hfil
\vbox to 0.5\vsize{
 \includegraphics{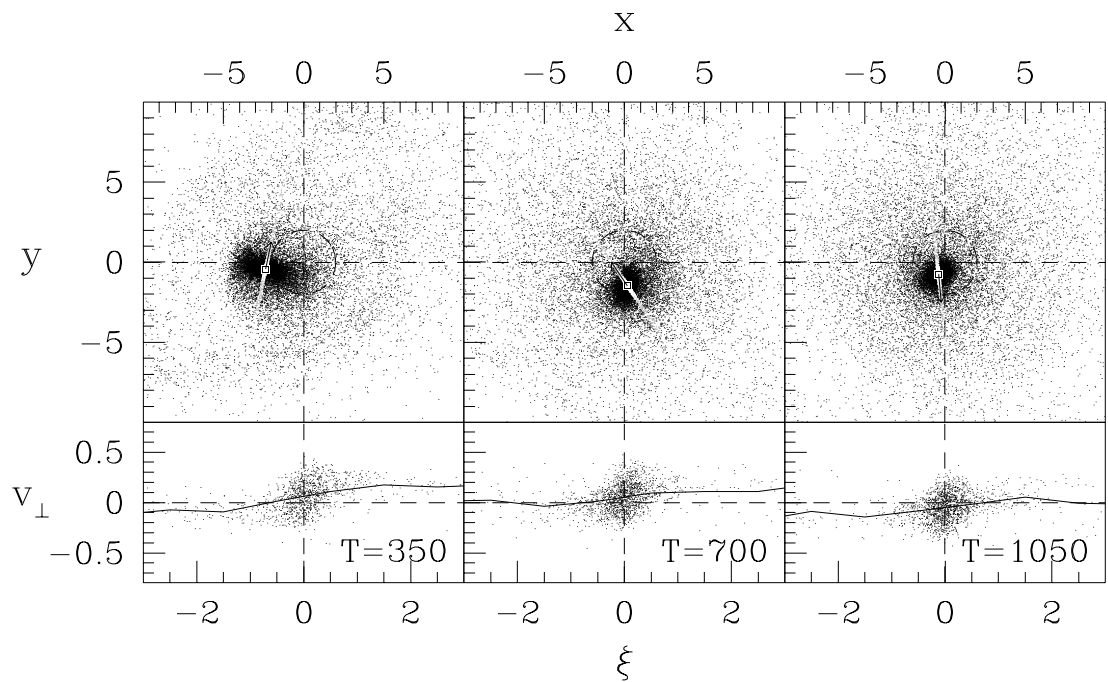}
}\hfil}
\figcaption{Upper: particle positions in a 3-D simulation in which the
disk started with spin retrograde to its orbit around the halo, and
its center at $x_0 = 2.5$.  The disk particles form a bar which is
displaced from the halo center, at the origin; the dashed line marks
the halo core $r_c = 2$.  Lower: particle velocity normal to a `slit'
along the bar minor axis, shown in the upper frames.  Distance is
measured from the centroid of the most-bound particles (marked by an
open square), in the direction of the arrow.}
\end{figure}

\newpage

\begin{figure}
\hbox to \hsize{\hfil
\vbox to 0.5\vsize{
 \includegraphics{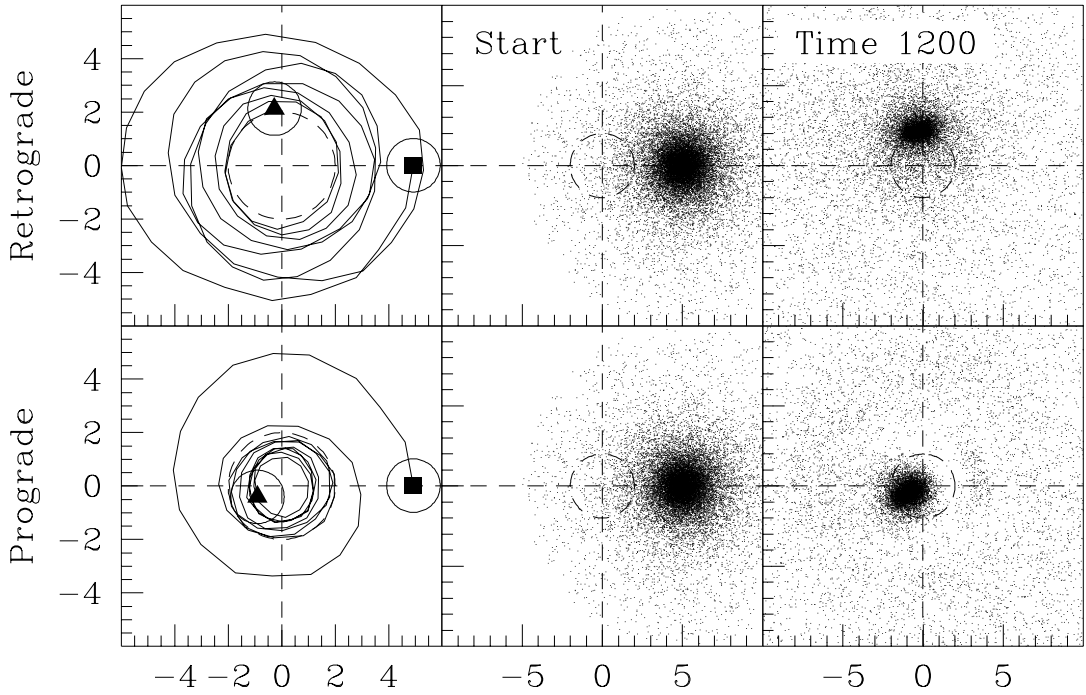}
}\hfil}
\figcaption{Path of the most tightly-bound particles, from the start
(square) to the finish (triangle) of a 2-D simulation beginning with
the disk center at $x_0 = 5$: the solid circles show a radius of 1
disk scale length, the dashed circle marks the halo core radius $r_c =
2$.}
\end{figure}

\end{document}